# Crystal structure and Raman active lattice vibrations of magnetic topological insulators MnBi$_2$Te$_4$·$n$(Bi$_2$Te$_3$) ($n$ = 0, 1, . . . , 6)

I. R. Amiraslanov,[1,2,*] Z. S. Aliev,[2,1] P. A. Askerova,[1] E. H. Alizade,[1] Y. N. Aliyeva,[2,1] N. A. Abdullayev,[1,2] Z. A. Jahangirli,[1,2] M. M. Otrokov,[3,4] N.T. Mamedov,[1,2,†] and E. V. Chulkov[5,6]

[1]*Institute of Physics, Azerbaijan National Academy of Sciences, Baku AZ1143, Azerbaijan*
[2]*Baku State University, Baku AZ1148, Azerbaijan*
[3]*Centro de Fisica de Materiales (CFM-MPC), Centro Mixto CSIC-UPV/EHU, Donostia-San Sebastian, Basque Country, Spain*
[4]*IKERBASQUE, Basque Foundation for Science, Bilbao, Spain*
[5]*Departamento de Polímeros y Materiales Avanzados: Física, Química y Tecnología,
Facultad de Ciencias Químicas, Universidad del País Vasco UPV/EHU,
20080 Donostia-San Sebastián, Basque Country, Spain*
[6]*Donostia International Physics Center (DIPC),
20018 Donostia-San Sebastián, Basque Country, Spain*

(Dated: July 20, 2022)

Further to the structure of the intrinsic magnetic topological insulators MnBi$_2$Te$_4$ $n$(Bi$_2$Te$_3$) with $n$<4, where index $n$ is the number of quintuple Te-Bi-Te-Bi-Te building blocks inserted between the neighboring septuple Te-Bi-Te-Mn-Te-Bi-Te building blocks, the structure of the members with $n$=4, 5 and 6 was studied using X-ray powder diffraction. The unit cell parameters and atomic positions were calculated. The obtained and available structural data were summarized to show that the crystal structure of all members of MnBi$_2$Te$_4$· $n$(Bi$_2$Te$_3$) follows the cubic close packing principle, independently of the space group of the given member. Confocal Raman spectroscopy was then applied. Comparative analysis of the number, frequency, symmetry, and broadening of the vibration modes responsible for the lines in the Raman spectra of the systems with $n$=1,. . . ,6, as well as MnBi$_2$Te$_4$ ($n$=0) and Bi$_2$Te$_3$ ($n$= ∞) has shown that lattice dynamics of MnBi$_2$Te$_4$ $n$(Bi$_2$Te$_3$) with $n$>0 overwhelmingly dominates by the cooperative atomic displacements in the quintuple building blocks.

## I. INTRODUCTION

The members of the tetradymite homological series MnBi$_2$Te$_4$ ·$n$(Bi$_2$Te$_3$) are layered systems whose structure consists of quintuple Te-Bi-Te-Bi-Te (QL) and septupleTe-Bi-Te-Mn-Te-Bi-Te (SL) layer building blocks, sepa-rated by van der Waals (vdW) gaps [1]. These systemsare currently considered as natural intrinsic magnetic topological heterostructures with index $n$ indicating the number of non-magnetic QLs inserted between the neighboring magnetic SLs [2]. Along with such prerequisites of non-trivial topology as strong spin-orbit interaction and inverted bandgap [3-5], the distinctive feature of these quantum materials is a magnetic gap that appears within topological surface states, providing necessary conditions for the realization of quantum anomalous Hall state [6-8].

Although existence of the members with $n$ up to 6 has been reported [2], in fact only the members with $n$=0,1,2 and 3 have been studied in detail using various experimental and theoretical techniques [1, 2, 9-44]. In particular, available structure refinement and Raman scattering studies [1, 26-44] have been performed for systems with $n$ less than 4, probably, because of the obvious difficulties in the preparation of MnBi$_2$Te$_4$ $n$(Bi$_2$Te$_3$) with high index $n$ and unavailability of samples.

We have succeeded in this regard and present here the crystal structure of the members with $n$=4, 5 and 6, summarize all available structural data for MnBi$_2$Te$_4$ ·$n$(Bi$_2$Te$_3$), trace down the evolution of Raman spectra of MnBi$_2$Te$_4$· $n$(Bi$_2$Te$_3$) that embraces all $n$ from 0 to 6 as well as $n$∞ (Bi$_2$Te$_3$), and provide a thorough analysis of the obtained Raman data.

## II. SYNTHESIS AND CRYSTAL STRUCTURE OF MnBi$_2$Te$_4$·$n$(Bi$_2$Te$_3$)

### A. Synthesis and crystal growth

Prior to the growth of the title compounds, the binaries MnTe and Bi$_2$Te$_3$ as precursors were synthesized from high purity (99.999 wt. %) elemental manganese, bismuth and tellurium purchased from Alfa Aesar. The synthesis of MnTe and Bi$_2$Te$_3$ was carried out in sealed quartz ampoules by melting the above elements at 1180 °C and 630 °C, respectively. To avoid the reaction of manganese with silica during melting, the inner wall of the ampoule was coated with graphite by thermal decomposition of acetone. The polycrystalline MnBi$_2$Te$_4$ $n$(Bi$_2$Te$_3$) compounds with different $n$ were synthesized by co-melting binaries in evacuated silica ampoules. The temperature of synthesis was varied depending on the required $n$. For example, for MnBi$_2$Te$_4$ ($n$=0) this temperature was 980 °C at which the melt was kept for 8 hours, while MnBi$_4$Te$_7$ ($n$=1) was synthe-sized at 900 °C, keeping the melt at this temperature for 8 hours. The single-crystalline ingots of the title compounds were then grown from pre-synthesized polycrystalline alloys using the vertical Bridgman-Stockbarger method described elsewhere [1].

---

* iamiraslan@gmail.com
† n.mamedov.physics@bsu.edu.az

## B. X-ray diffraction characterization

X-ray diffraction (XRD) patterns were obtained using a BRUKER XRD D2 Phaser (Cu$K_\alpha$; $5 \leq 2\theta \leq 100°$) diffractometer. The Rietveld refinements of the crystal structures were performed using powder diffraction data. The phase constitution analyses and structural refinements were performed using the EVA and TOPAS-4.2 software packages supplied together with the diffractometer.

The single-crystalline samples with an average size of 2x2x0.1 mm$^3$ were cleaved from the as-grown ingots of each homologous phase by the top-off procedure under an optical microscope. The (0001) XRD patterns of these samples are given in Fig. 1.

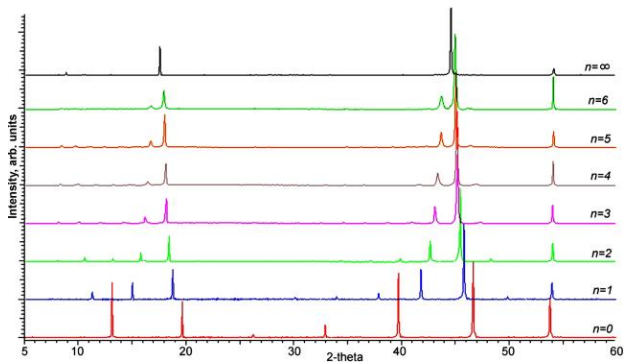

FIG. 1. The (0001) diffraction patterns of MnBi$_2$Te$_4$·$n$(Bi$_2$Te$_3$) with different index $n$

The space group and lattice parameters for each of the obtained members of the $n$MnBi$_2$Te$_4$ $n$(Bi$_2$Te$_3$) are given in Table I. Note that along with the data previously published [1, 2] and reproduced here for structures with n= $\infty$, 1, 2 and 3, this table also shows the data extracted in this work for n=4, 5 and 6.

The details regarding the newly extracted data are given in the next two subsections, together with the general building principle of MnBi$_2$Te$_4$·$n$(Bi$_2$Te$_3$) structures.

## C. Rietveld refinement of the crystal structure of MnBi$_{10}$Te$_{16}$ ($n$=4)

Note that the atomic percent of manganese in MnBi$_2$Te$_4$·$n$(Bi$_2$Te$_3$) is decreasing as 100/(5$n$+7) with increasing $n$ due to the stably increasing number of the Bi$_2$Te$_3$ QLs. Therefore, conduction of the conventional Rietveld refinement of the structures with large $n$ requires sufficiently massive single crystals. While high $n$ tiny single crystals of a smaller volume can be successfully obtained from the as-grown ingots by cleaving and have already been used in reported XRD studies [1] and Raman studies presented in this work, the Rietveld refinement of the crystal structure for $n>3$ has not been performed yet, because of the absence of enough massive single-phase crystalline samples.

In this work, we were unable to find a massive single-phase sample among the pieces detached from the as-grown nominally $n$=4 ingot either. Nevertheless, a sample that contained a mixture of single-crystalline $n$=4 (50.7%) and $n$=3 (49.3%) phases and was suitable for structural analysis has been available.

Figure 2 displays the XRD pattern of this sample, together with Rietveld fitted pattern. A very good agreement between both patterns is observed. The space group and unit cell parameters obtained using Le Bail's refinement for $n$=4 phase have been given earlier in Table I. Interatomic Mn-Te and Bi-Te distances are 2.98(3) and 2.94(4)-3.34(4)Å, respectively. Atomic positions refined up to $R_{Bragg}$=2.34 % are given in supplementary (Table S1), together with similar data for the other members of the MnBi$_2$Te$_4$·$n$(Bi$_2$Te$_3$) homologous series.

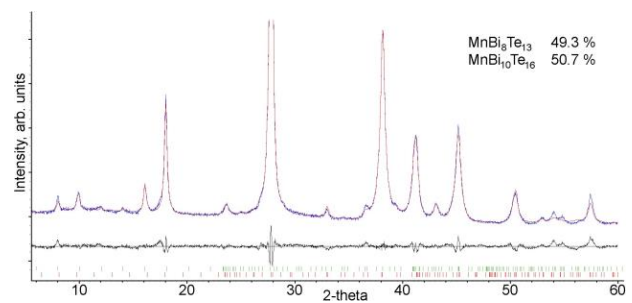

FIG. 2. XRD (blue curve) and Rietveld fitted (red curve) patterns of a crystalline mixture of 49.3 % MnBi$_8$Te$_{13}$ and 50.7% MnBi$_{10}$Te$_{16}$. The difference curve is given beneath these patterns

## D. Crystal structure of MnBi$_2$Te$_4$·$n$(Bi$_2$Te$_3$) and cubic close packing principle

In layered crystals, the stacking manner between the building blocks separated by vdW gaps usually varies leading to different polymorphs (polytypic forms) with cubic or hexagonal close packing or just to stacking faults in such polymorphs. A weak vdW bond provides favorable conditions for the above blocks to easily shift relative to each other overcoming small enough small energy barriers between the various stable states. GaSe and MoS$_2$, which have a lot of polymorphs [45, 46] are just few examples to mention among the vast number of layered crystals.

TABLE I. The space group and lattice parameters of MnBi$_2$Te$_4 \cdot n$(Bi$_2$Te$_3$) with different $n$

| MnBi2Te4·$n$(Bi2Te3) | | $n$ | Space group | Unit cell parameters | | QL (5) and SL (7) sequences |
|---|---|---|---|---|---|---|
| Stoichiometmetric formula | Unit cell content | | | $a$ (Å) | $c$ (Å) | |
| Bi$_2$Te$_3$ | Bi$_2$Te$_3 \times 3$ | ∞ | $R$-3$m$ | 4.386 | 30.497 | -5- |
| MnBi$_2$Te$_4$ | MnBi$_2$Te$_4 \times 3$ | 0 | $R$-3$m$ | 4.3304(1) | 40.956(2) | -7- |
| MnBi$_4$Te$_7$ | MnBi$_4$Te$_7$ | 1 | $P$-3$m$1 | 4.3601(1) | 23.798(2) | -5-7- |
| MnBi$_6$Te$_{10}$ | MnBi$_6$Te$_{10} \times 3$ | 2 | $R$-3$m$ | 4.3685(2) | 101.870(7) | -5-5-7- |
| MnBi$_8$Te$_{13}$ | MnBi$_8$Te$_{13} \times 3$ | 3 | $R$-3$m$ | 4.3927(8) | 132.336(24) | -5-5-5-7- |
| MnBi$_{10}$Te$_{16}$ | MnBi$_{10}$Te$_{16}$ | 4 | $P$-3$m$1 | 4.3701(7) | 54.304(9) | -5-5-5-5-7 |
| MnBi$_{12}$Te$_{19}$[a] | MnBi$_{12}$Te$_{19} \times 3$ | 5 | $R$-3$m$ | 4.370 | 193.50 | -5-5-5-5-5-7- |
| MnBi$_{14}$Te$_{22}$[a] | MnBi$_{14}$Te$_{22} \times 3$ | 6 | $R$-3$m$ | 4.370 | 223.89 | -5-5-5-5-5-5-7- |

[a] For structures with n=5 and 6, parameter a is taken equal to that of the structure with n=4, while parameter $c$ is calculated from the diffraction patterns (Fig. 1, n=5; 6).

As a matter of fact, all MnBi$_2$Te$_4$ $n$(Bi$_2$Te$_3$) structures are built up only according to the cubic close packing rule (Fig. 3), even if the space group of the structure changes from $R$-3$m$ for the members with $n$=∞, 0, 2, and 3 to $P$-3$m$1 for $n$=1 and 4 (Table I). The latter fact is illustrated by Fig. 3 with $n$=3 ($R$-3$m$) and $n$=4 ($P$-3$m$1) structures shown for comparison.

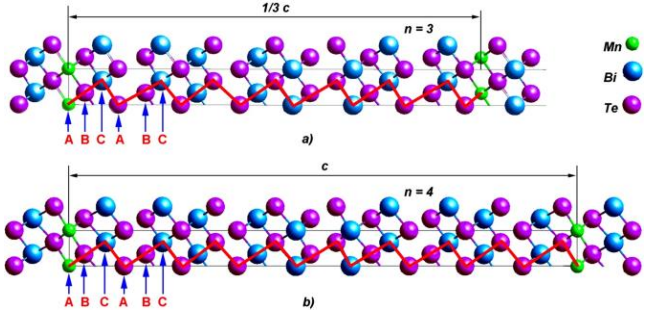

FIG. 3. Crystal structure of MnBi$_8$Te$_{13}$ with space group $R$-3$m$ (a) and MnB$_{10}$Te$_{16}$ with space group $P$-3$m$1(b). A, B and C indicate closely packed atoms with coordinates (0, 0), (1/3, 2/3) and (2/3, 1/3) in $ab$ plane. Red lines guide close packing (. . . ABC. . . ) of Mn, Bi and Te along the $c$ direction.

Thus, until now only the rhombohedral ($R$-type) unit cell is found for MnBi$_2$Te$_4$ $n$(Bi$_2$Te$_3$) with $n$= ∞, 0, 2, and 3, while the $P$-type trigonal or hexagonal polymorphs are absent. On the other hand, the $R$-type polymorph is not found for $n$=1 and $n$=4 members of the MnBi$_2$Te$_4$ $n$(Bi$_2$Te$_3$) family which crystallize into a trigonal structure with primitive ($P$-type) unit cell. Extending this rule to higher $n$ members it is easy to predict their structures and space groups. The latter can be only $P$-3$m$1 if one-third of the number of atoms in a chemical formula is an integer or $R$-3$m$ in the rest cases. Moreover, following this rule and using the already accurately known structural data of SLs and QLs from which each MnBi$_2$Te$_4 \cdot n$(Bi$_2$Te$_3$) structure is built up, it is not difficult to calculate the lattice parameters and atomic positions for arbitrary $n$. It is exactly in this way the structure of the members with $n$=5 and $n$=6 has been described after a comparison of the calculated data with the results of the XRD examination. Tables S1 and S2 provide atomic coordinates for all MnBi$_2$Te$_4$ $n$(Bi$_2$Te$_3$) studied in this work.

### III. RAMAN SPECTRA OF MnBi$_2$Te$_4 \cdot n$(Bi$_2$Te$_3$)

#### A. Experimental technique

Raman scattering of MnBi$_2$Te$_4 \cdot n$(Bi$_2$Te$_3$) was studied with the aid of a confocal micro-spectrometer Nanofinder 30 (Tokyo Instr., Japan). Second harmonic (532 nm) of Nd:YAG laser with a maximum output power of 10 mW was used as excitation source, and the cross-sectional beam diameter was 4 $\mu$m. Diffraction grating with 1800 grooves per mm provided a spectral resolution of 0.5 cm$^{-1}$. The spectra were detected using photon counting CCD camera cooled down to −100 °C, the exposure time was usually 1 min. The Raman signal was filtered out using edge filters LP03-532RU-50 (Semrock Company). The study was conducted in backscattering geometry.

#### B. Raman active modes of MnBi$_2$Te$_4 \cdot n$(Bi$_2$Te$_3$)

According to the performed site-symmetry analysis based on the structural data in Tables S1 and S2, the number of vibration modes active in Raman scattering of MnBi$_2$Te$_4$ $n$(Bi$_2$Te$_3$) has to grow with rising $n$ from 0 to 6 (Table II, third row), because of the growth of the number of atoms in the elementary cell.

As shown in Fig. 4 with normalized experimental spectra (black lines) and deconvoluted spectra (green lines) of MnBi$_2$Te$_4$ $\cdot n$(Bi$_2$Te$_3$), in fact, the number of these modes is not following above expectation. While the number of lines in Raman spectra of the terminate members, MnBi$_2$Te$_4$ ($n$=0) and Bi$_2$Te$_3$ ($n=\infty$) corresponds to the number of lines given by the site-symmetry analysis for bulk crystals, this number for the rest members of MnBi$_2$Te$_4$ $n$(Bi$_2$Te$_3$) is constant and equals 4, independently of $n$ value. Not shown in Fig. 4, the mode $E_g^1$ of MnBi$_2$Te$_4$ ($n$=0) has been experimentally detected in a couple of recent works [41, 43], making the total number of the observed lines in the Raman spectrum equal to 6. Disappearance of the mode $E_g^1$ from Raman spectra of MnBi$_2$Te$_4$ $n$(Bi$_2$Te$_3$) with $n$=5, 6 and $\infty$ in Fig. 4 is caused by instrumental limitations and the total number (4) of lines is preserved.

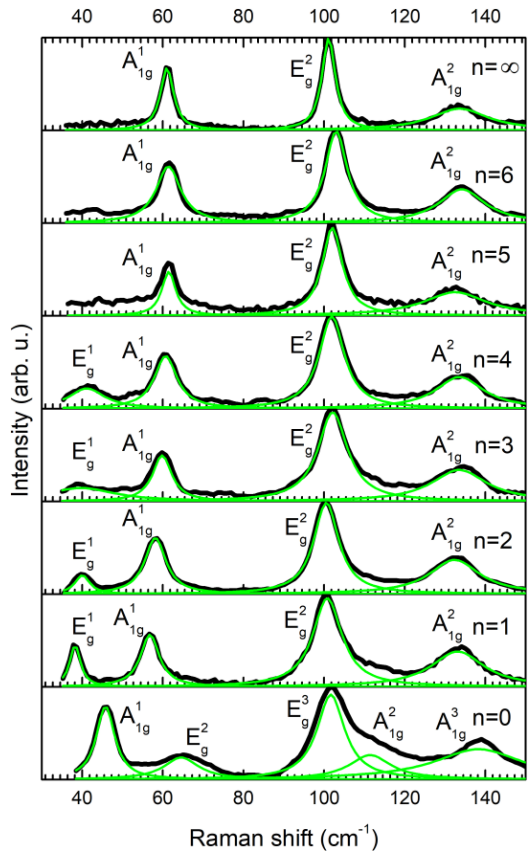

FIG. 4. Normalized Raman spectra (black lines) of MnBi$_2$Te$_4$ $n$(Bi$_2$Te$_3$) for n=0,...,6 and $\infty$ The results of deconvolution are shown by green lines.

For example, MnBi$_4$Te$_7$ ($n$=1) shows only 4 (Fig. 4) out of the expected 12 modes (Table II). For the rest members of MnBi$_2$Te$_4$ $n$(Bi$_2$Te$_3$) with $1<n<\infty$ the difference between the expected and real number of lines is even more essential by comparison.

The line-width given in Table III for each mode observed in Raman spectra in Fig.4 witnesses the absence of any broadening of the Raman lines. Such broadening would have manifested itself if the number of the Raman lines were increasing with rising $n$. In other words, Raman modes in Fig. 4 are the only normal vibrations active in Raman scattering of MnBi$_2$Te$_4 \cdot n$(Bi$_2$Te$_3$).

For MnBi$_2$Te$_4$ ($n$=0) that was studied thoroughly [1,30,38-44] and consists only from SLs, symmetry notations of vibration modes in Fig. 4 coincide with those of recent work [41] where these symmetries were confirmed by the measurements of the polarization of the modes. Starting from MnBi$_4$Te$_7$ ($n$=1), for all members containing both SLs and QLs, and for Bi$_2$Te$_3$ ($n\infty$), the symmetry of each mode in Fig. 4 follows the theoretical work [47] that considered Raman active modes of free-standing QLs of Bi$_2$Te$_3$. The reason for this is a very close correspondence between the results calculated for the $n$ number of QLs of Bi$_2$Te$_3$ and obtained for MnBi$_2$Te$_4$ $n$(Bi$_2$Te$_3$) with the same index $n$. Such correspondence becomes obvious after the careful comparison of the obtained Raman data and the available results for Bi$_2$Te$_3$.

### C. Lattice dynamics of MnBi$_2$Te$_4 \cdot n$(Bi$_2$Te$_3$) and Bi$_2$Te$_3$

The frequency behavior of the Raman active modes with changing $n$ in MnBi$_2$Te$_4 \cdot n$(Bi$_2$Te$_3$) is shown in Fig. 5.

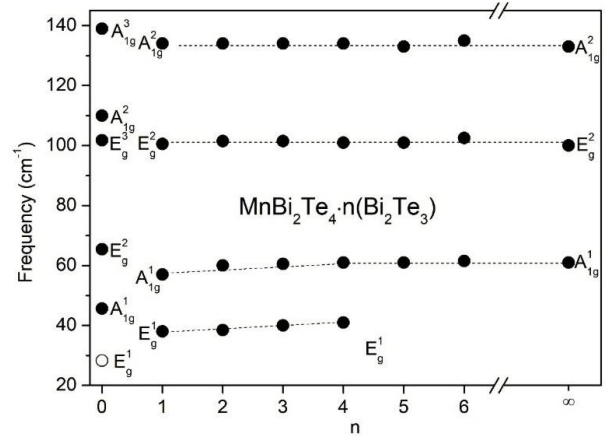

FIG. 5. The dependence of the frequency of the modes observed in the Raman spectra on the number ($n$) of QLs inserted between SLs in MnBi$_2$Te$_4$ $n$(Bi$_2$Te$_3$). Dashed lines connect the frequencies of the modes with similar displacement fields. The open circle in Fig. 5 indicates the reported value [41, 43] of the frequency of $E_g^1$ mode of MnBi$_2$Te$_4$ ($n$=0). The modes with the same symmetry for $n$=0 (MnBi$_2$Te$_4$) and $n$=1 (MnBi$_4$Te$_7$) are not connected because of the different displacement fields, as explained in the text.

The most noticeable frequency change occurs within the first two members of MnBi$_2$Te$_4$ $n$(Bi$_2$Te$_3$) with $n$=0 and $n$=1. As seen from Fig. 5, except for those of $E_g^2$ and $E_g^3$ the frequencies of the modes with the same symmetry in MnBi$_2$Te$_4$ and MnBi$_4$Te$_7$ noticeably differ from each other in a similar frequency range. (Note that in spite of the same symmetry the displacement fields of the modes are different.). In Fig. 5 the open circle indicates the reported value [41, 43] of the frequency of $E_g^1$ mode of MnBi$_2$Te$_4$ ($n$=0). This mode corresponds to the in-plane in-phase displacements of Bi and Te in SLs. The other in-plane in-phase mode of MnBi$_2$Te$_4$ is $A_{1g}^1$. The frequencies of the in-plane in-phase counterparts of these modes in MnBi$_4$Te$_7$ ($n$=1) are appreciably higher and obey the relation $\omega_{n=1}/\omega_{n=0} \approx \left[\frac{(2m_{Te}+m_{Bi})}{(m_{Te}+m_{Bi})}\right]^{1/2}$, where $m_{Te}$ and $m_{Bi}$ are the atomic masses of Te and Bi. It is then suggestive to ascribe the low-frequency modes of MnBi$_4$Te$_7$, $E_g^1$ and $A_{1g}^1$ to the displacements of Te and Bi in QLs of Bi$_2$Te$_3$ rather than SLs like in the case of MnBi$_2$Te$_4$. Contrary to the expectations, the number of the modes observed after the insertion of just one QL of Bi$_2$Te$_3$ between SLs of MnBi$_2$Te$_4$ to form MnBi$_4$Te$_7$ is not increasing but is down to the number of the modes observed for Bi$_2$Te$_3$ (Table II). Importantly enough, the number of modes and their symmetry for one and more QLs of Bi$_2$Te$_3$ and bulk Bi$_2$Te$_3$ is the same and the low-frequency modes of one QL of Bi$_2$Te$_3$ are red shifted in comparison with their bulk counterparts [47]. According to Fig. 5, quite similar dispersion is observed for the low-frequency $E_g^1$ and $A_{1g}^1$ modes in structures with $n$>0.

TABLE II. The number (N) of atoms in the unit cell (third row), the number of modes of $A_{1g}$ and $E_g$ symmetry (fourth row), and the total number of modes (fifth row) for each member of MnBi$_2$Te$_4 \cdot n$(Bi$_2$Te$_3$). $E_g$ mode is doubly degenerate

| Compound | Bi$_2$Te$_3$ | MnBi$_2$Te$_4$ | MnBi$_4$Te$_7$ | MnBi$_6$Te$_{10}$ | MnBi$_8$Te$_{13}$ | MnBi$_{10}$Te$_{16}$ | MnBi$_{12}$Te$_{19}$ | MnBi$_{14}$Te$_{22}$ |
|---|---|---|---|---|---|---|---|---|
| $n$ | $\infty$ | 0 | 1 | 2 | 3 | 4 | 5 | 6 |
| $N$ | 5 | 7 | 12 | 17 | 22 | 27 | 32 | 37 |
| Modes | $2A_{1g} + 2E_g$ | $3A_{1g} + 3E_g$ | $6A_{1g} + 6E_g$ | $8A_{1g} + 8E_g$ | $11A_{1g} + 11E_g$ | $13A_{1g} + 13E_g$ | $16A_{1g} + 16E_g$ | $18A_{1g} + 18E_g$ |
| Number | 4 | 6 | 12 | 16 | 22 | 26 | 32 | 36 |

TABLE III. Full width at half maximum (FWHM) of the Raman lines of the vibration modes of MnBi$_2$Te$_4$ $n$(Bi$_2$Te$_3$) with different n

| $n$ | $E_g^1$ | $E_g^2$ | $E_g^3$ | $A_{1g}^1$ | $A_{1g}^2$ | $A_{1g}^3$ |
|---|---|---|---|---|---|---|
| | | FWHM (cm$^{-1}$) | | | | |
| 0 | - | 10.5 | 8.3 | 5.54 | 12.6 | 26.6 |
| 1 | 3.2 | 8.7 | - | 5.7 | 15.8 | |
| 2 | 4.5 | 7.5 | - | 6.4 | 15.2 | |
| 3 | 21 | 8.9 | - | 5.8 | 16.5 | |
| 4 | 10 | 7.6 | - | 6.8 | 14.4 | |
| 5 | - | 6.1 | - | 4.2 | 19.1 | |
| 6 | - | 6.7 | - | 7.4 | 13.9 | |
| $\infty$ | - | 4 | - | 4.0 | 17.4 | |

In other words, there has to be a one-to-one correspondence between our results for structures with $n$=1,2,3,4,5 and 6 and those for one, two, three, four, five and six QLs of Bi$_2$Te$_3$. In fact, our results practically coincide with the results obtained by Zhao et. al. [48] for 2-11 QLs of Bi$_2$Te$_3$ on SiO$_2$/Si substrate. According to the later results, the low-frequency $E_g^1$ appeared only for 2 QLs of Bi$_2$Te$_3$ and was no longer observed for greater number of QLs, which corroborates the dying intensity of this mode with increasing index n in our case. Another coincidence is the blue shift of $A_{1g}^1$ mode with rising the number of QLs from 2 to 4, as well as the fixed frequency of this mode for larger number of QLs [48].

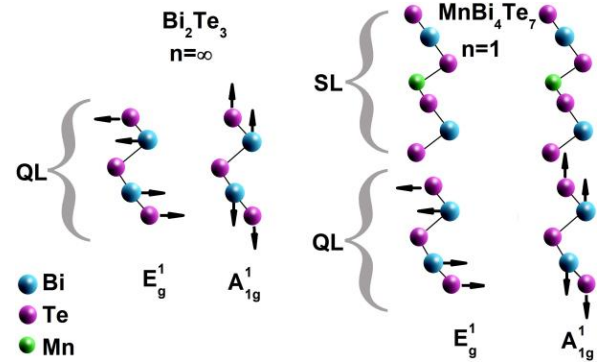

FIG. 6. On the left: displacement fields of the low-frequency Raman active modes $E_g^1$ and $A_{1g}^1$ of Bi$_2$Te$_3$ ($n$=$\infty$). On the right: displacement fields of $E_g^1$ and $A_{1g}^1$ Raman active modes of MnBi$_4$Te$_7$ ($n$=1) - right figure. The arrows indicate only the direction of the displacements occurring parallel (horizontal arrows) and perpendicular (vertical arrows) to the layer plane for $E_g^1$ and $A_{1g}^1$ type vibrations, respectively.

Above described drives to a conclusion that the displacement fields of the low-frequency lattice modes of MnBi$_2$Te$_4$ $n$(Bi$_2$Te$_3$) with $n$=1 and higher are as shown in Fig. 7 with displacement fields of these modes in Bi$_2$Te$_3$ (on the left) and MnBi$_4$Te$_7$ (on the right).

The high-frequency anti-phase out-of-plane modes, $E_g^2$ and $A_{1g}^2$ for $n$>0 in Fig. 5 are difficult to attribute to QLs or SLs or to both simultaneously. The frequencies of $E_g^2$ modes of MnBi$_2$Te$_4$ and the rest members of MnBi$_2$Te$_4$ ·$n$(Bi$_2$Te$_3$) are identical and the modes are indistinguishable from one another. The $A_{1g}^2$ mode of MnBi$_2$Te$_4$ no longer exists for $n$>0 and the closest to the $A_{1g}^2$ mode of the structures with $n$>0 is $A_{1g}^3$ mode of MnBi$_2$Te$_4$ (Fig. 5). Although there is a small (2%) frequency difference between $A_{1g}^3$ mode of the structure with $n$=0 and $A_{1g}^2$ mode of the structures with $n$>0, change, this is not enough to attribute $A_{1g}^2$ mode to QLs. Decisive in this regard (it concerns the $E_g^2$ mode, too) is the fact that participation of QLs in low-frequency vibrations dictates participation of QLs in high-frequency vibrations, too. Such assignment is strongly supported by the limited fixed-to-four number of the Raman lines observed for MnBi$_2$Te$_4$·$n$(Bi$_2$Te$_3$) with $n>0$.

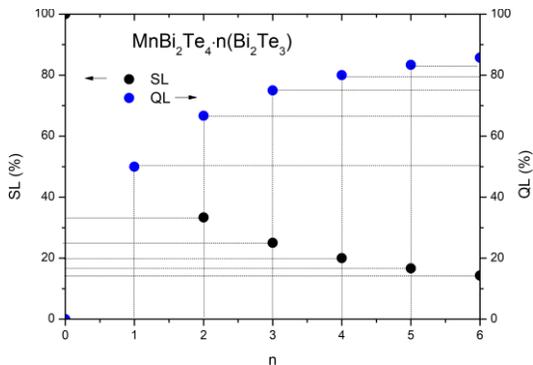

FIG. 7. Percentage of SLs (left scale, black solid circles) and QLs (right scale, blue solid circles) in MnBi$_2$Te$_4$ $n$(Bi$_2$Te$_3$), taken as 100% /($n$+1) and 100% $n$/($n$+1) for SLs and QLs, respectively.

As seen in Fig.7, while the share of QLs in the scattering volume increases, the share of SLs decreases, reaching only 14%. However, even this not large share (not speaking of 50% for $n$=1) is quite enough to detect Raman modes of SLs by such a sensitive technique as confocal Raman spectroscopy, if such modes would have existed and manifested themselves like in the case of even one SL of MnBi$_2$Te$_4$ [43].

Thus, Raman-active part of lattice dynamics of MnBi$_2$Te$_4$·$n$(Bi$_2$Te$_3$) with n in the range $\leq 1 \leq \infty$ is determined by lattice dynamics of $n$QLs of Bi$_2$Te$_3$. Note that this is not too surprising since such dynamics are a prerequisite of the limiting case of Bi$_2$Te$_3$.

tem with $n$=4 by Rietveld refinement and calculated for the systems with $n$=4 and 5 by using the cubic close packing rule and their XRD patterns. This principle is shown to underlie the crystal structure arrangement of each member of MnBi$_2$Te$_4$ $n$(Bi$_2$Te$_3$), accounts for the absence of polymorphs, and permits only $R$-3$m$ or $P$-3$m$1 space group of the members of this homological series. Atomic positions, space groups and lattice parameters are summarized for all available MnBi$_2$Te$_4$ $n$(Bi$_2$Te$_3$) to provide a database demanded for theoretical calculations.

Raman spectra, earlier available only for the systems with index $n<4$, have been obtained for systems with $n$=4,5 and 6. Essential disagreement between the results of site-symmetry analysis and experimental observations regarding the number of the vibration modes active in Raman scattering of MnBi$_2$Te$_4$ $n$(Bi$_2$Te$_3$) with index $n$ in the range $1 \leq n \ll \infty$ has been found. This disagreement can be lifted off if the lattice dynamics of MnBi$_2$Te$_4$ $n$(Bi$_2$Te$_3$) with given index n and that of $n$QLs of Bi$_2$Te$_3$ are similar. The only difference between the MnBi$_2$Te$_4$ $n$(Bi$_2$Te$_3$) and Bi$_2$Te$_3$ is that displacement field of each mode in the former includes atomic displacements in QLs and infinitesimally small atomic displacements in SLs, while only atomic displacements in QLs form lattice modes in the latter.

Magnetic Mn atoms do not participate in displacement fields of the modes active in Raman scattering but contribute in IR -active modes which are yet to be studied. Besides, MnBi$_2$Te$_4$·$n$(Bi$_2$Te$_3$) systems with high index $n$ are of special interest, because of the crossover between the 3D ($n$=3[49]) and 2D ferromagnetic cases. The system with $n$=4 is bordering these cases while systems with $n$=5 and 6 are very likely to have truly magnetically decoupled SLs, bringing about qualitatively new conditions for the realization of quantum anomalous Hall state.

## IV. SUMMARY

MnBi$_2$Te$_4$ $n$(Bi$_2$Te$_3$) single crystals with integer index $n$ running from 0 to 6 have been successfully grown and characterized by X-ray powder diffraction (XRD) and confocal Raman spectroscopy.

The crystal structure has been established for the sys-

## ACKNOWLEDGMENTS

This work, carried out within the project NovelMTI, was supported by grant-in-aid from the Azerbaijan National Academy of Sciences. M.M.O. acknowledges the support by Spanish Ministerio de Ciencia e Innovación (Grant No. PID2019-103910GB-100).